\documentclass[aps,prc,reprint,unsortedaddress,superscriptaddress]{revtex4-2}
\usepackage[latin1]{inputenc}
\usepackage{hyperref}
\usepackage{amsmath}
\usepackage{amsfonts}
\usepackage{amssymb}
\usepackage{graphicx}
\usepackage{xcolor}
\usepackage[per-mode=reciprocal]{siunitx}
\usepackage{multirow}

\usepackage{booktabs} %for \cmidrule

\begin{document}

%%%%%%%%%%%%%%%%%%%%%%%%%%%%%%%%%%%%%%%%%%%%%%%%%%%%%%%%%%%%%%%%%%%%%%%%%%%%%%%%%%%%%%%%%%%%%%%%%%%%%%%%%%%%
%\title{ Testing the Young-Koppel Model with ultracold neutrons}

%\title{ Ultracold neutron transmission of D$_2$ gas as a test of the Young-Koppel model}
\title{Temperature dependent ultracold neutron transmission in D$_2$ gas $-$ a test of the Young-Koppel model}

\author{G.~Bison}
\affiliation{Laboratory for Particle Physics, PSI Center for Neutron and Muon Sciences, Paul Scherrer Institute, Forschungsstrasse 111, 5232 Villigen PSI, Switzerland}
\author{R.~Gr\"ossle}
\email[E-mail:]{robin.groessle@kit.edu}
\affiliation{KIT - Karlsruhe Institute of Technology, DE-76187 Karlsruhe, Germany}
\author{K.~Kirch}
\affiliation{Laboratory for Particle Physics, PSI Center for Neutron and Muon Sciences, Paul Scherrer Institute, Forschungsstrasse 111, 5232 Villigen PSI, Switzerland}
\affiliation{Institute for Particle Physics and Astrophysics, ETHZ, 8093 Zurich, Switzerland} 
\author{B.~Lauss}
\affiliation{Laboratory for Particle Physics, PSI Center for Neutron and Muon Sciences, Paul Scherrer Institute,  Forschungsstrasse 111, 5232 Villigen PSI, Switzerland}
\author{F.~Priester}
\affiliation{KIT - Karlsruhe Institute of Technology, DE-76187 Karlsruhe, Germany}
\author{I.~Rien\"acker}
\affiliation{Laboratory for Particle Physics, PSI Center for Neutron and Muon Sciences, Paul Scherrer Institute, Forschungsstrasse 111, 5232 Villigen PSI, Switzerland}
\author{G.~Zsigmond}
\email[E-mail:]{geza.zsigmond@psi.ch}
\affiliation{Laboratory for Particle Physics, PSI Center for Neutron and Muon Sciences, Paul Scherrer Institute, Forschungsstrasse 111, 5232 Villigen PSI, Switzerland}

%%%%%%%%%%%%%%%%%%%%%%%%%%%%%%%%%%%%%%%%%%%%%%%%%%%%%%%%%%%%%%%%%%%%%%%%%%%%%%%%%%%%%%%%%%%%%%%%%%%%%%%%%%%%

%%%%%%%%%%%%%%%%%%%%%%%%%%%%%%%%%%%%%%%%%%%%%%%%%%%%%%%%%%%%%%%%%%%%%%%%%%%%%%%%%%%%%%%%%%%%%%%%%%%%%%%%%%%%
\begin{abstract}

The Young-Koppel model (YK) describes 
%with high accuracy 
comprehensively
the interaction of slow neutrons with diatomic gases such as H$_2$ and D$_2$. This paper reports on the first experimental results of ultracold neutron (UCN) scattering over a wide temperature range vindicating the YK model for gaseous D$_2$ and showing an important difference in the temperature dependence to a low-energy low-temperature approximation (LETA).  LETA is confirmed, however, to be valid for monoatomic gases such as Ne. 
Calculated cross sections for other noble gases were also confirmed for ultracold neutrons.
Finally, the total cross section of UCNs in H$_2$ gas was measured and analyzed applying the Young-Koppel model, however, in a more limited temperature range, confirming the theoretical prediction. 

%Buzzwords: UCN D2 and H2 scattering is a fundamental test of the interaction of elemental particle (UCN) with a composed fermionic (H2) or bosonic (D2) system.
\end{abstract}

\maketitle

\section{Introduction}
\label{sec:introduction}
%--> 
%{\bf Geza}

Ultracold neutrons (UCN) are free neutrons with kinetic energies below about 300 neV that experience total reflection under all angles of incidence from surfaces of suitable materials~\cite{Golub1991}.
The UCN source at the Paul Scherrer Institute~\cite{Bison2020,Bison2022,Lauss2021scipost} is used for various fundamental particle physics experiments, such as the search for an electric dipole moment of the neutron~\cite{Abel2020,Ayres2021n2EDM} and a measurement of the free neutron lifetime~\cite{Auler2024tauspect}.

%For UCN production, a spallation source is used and the fast neutrons produced are cooled down first in a heavy water tank and finally in a frozen deuterium crystal above the deuterium crystal is an area filled with gaseous deuterium at several Kelvins. For extraction, the UCNs have to pass cold deuterium gas. At the moment the production of UCNs is about a factor of four lower than the original models predicted  Since the UCN rate is the main benchmark parameter for UCN production, this needs an improvement and will have a direct positive impact on all the experiments performed with UCNs at the PSI.

The present work investigates the interaction between slow neutrons and individual deuterium molecules.
This interaction, by its fundamental nature, has intrigued researchers 
from the early days of neutron physics~\cite{Schwinger1937,Hamermesh1939,Hamermesh1946} and remains highly topical until today in nuclear physics, for example, in high precision measurements of the hadronic interaction~\cite{Grammer2015,Blyth2018} and in the design of neutron sources~\cite{Zanini2022,Dijulio2023}.
It also has important practical implications for the  non-standard operating 
conditions of the PSI UCN source when considerable D$_2$
vapor pressure is present above the solid D$_2$~\cite{Ries2016},
as UCN-D$_2$ scattering can lead to significant UCN losses via up-scattering
- i.e. neutrons acquire large enough energies to leave the
UCN energy range.

%In order to acquire an improved understanding of the UCN source, 

In this work we tested the Young-Koppel model~\cite{YoungKoppel1964} 
that can be applied to describe the interaction between UCNs and H$_2$ or D$_2$ gases. 
It is the first time that the Young-Koppel model 
was tested over a wide temperature range from about 100\,K to above 300\,K. 
%\textcolor{red}{
The temperature dependence is its unique signature for distinguishing it from the monoatomic approximation.
%}
%%  is it 80  or  100 K to start with - then also in the back adapt 
%This is mandatory to calibrate the UCN production for normal operation with solid D$_2$, and the UCN source and beamline optics to find possibilities to improve the UCN production rate.
%
% Besides the positive impact on all of the UCN experiments, 
%THis is the first time that the Young-Koppel model has never been tested over a wide temperature range and therefore this will be a first-of-kind measurement on its own.

\section{Theoretical description}
\label{sec:theory}

Young and Koppel (YK) calculated the neutron scattering cross sections from hydrogen and deuterium molecules by taking into consideration the spin correlation effects, 
as well as rotation and vibration of the molecules~\cite{YoungKoppel1964,KoppelYoung1966}. 
In this approach, they considered vibrations as harmonic and decoupled from rotation. Also, there is no interaction between the molecules which would be the case in high-pressure gases or in liquids.
The YK model can also be applied for liquids, however, for incident neutrons with energies less than roughly the solid state  Debye temperature it is not accurate~\cite{YoungKoppel1964}. When the YK model was published, it was supported only by a limited amount of experimental data 
of the total cross section of neutrons in hydrogen gas
as a function of incident energy~\cite{SquiresStewart1955scattering}.

Later, the hydrogen gas form factor was measured as a function of scattering angle with inelastic neutron scattering~\cite{HerwigSimmons1992inelastic}. An excellent match was obtained with the YK model predictions, however, only in a very limited parameter space, at 18\,K and one incident energy. 
The authors also reported a 20~\% mismatch between the fitted and the observed temperatures, which they explained by a temperature gradient in the measurement cell.

A modified model for high pressures, assuming an effective temperature instead of the observed one, was tested in Ref.~\cite{CorradiEtAl2004hydrogen}. 
The measurements reported in~\cite{HerwigSimmons1992inelastic} were completed 
in the same parameter space as reported in~\cite{GuariniEtAl2005hydrogen}, 
in which the double-differential cross section of molecular hydrogen was measured 
at room temperature (RT) and could be fitted by using the YK model. 

A significant time later, the total cross section of deuterium gas at 25\,K  
was measured~\cite{AtchisonEtAl2005deuterium}.
However, the YK model and a low energy approximation model could not be distinguished at this single low temperature, thus it did not provide a full corroboration of the YK model. 
New gas measurements were performed at room temperature~\cite{SeestromEtAl2015upscattering,SeestromEtAl2017upscattering},
where the authors scaled with $\sqrt{T}$ the 25\,K-value 
reported in Ref.~\cite{AtchisonEtAl2005deuterium} to an RT value. 
With simple reasoning, in the coordinate system of the molecules, the UCNs propagate with the mean velocity of the Maxwell-Boltzmann gas, $w \propto \sqrt{T}$, and the interaction rate $\sigma w$ is invariant to changing the coordinate system (see also Eq.~(4) in Ref.~\cite{AtchisonEtAl2005deuterium}).
This assumption of the low-energy low-temperature 
approximation (LETA) is over-simplified at higher D$_2$ gas temperatures.
The LETA model is similar to the monoatomic gas model~\cite{Zemach1956PhysRev}, however, it uses an effective scattering cross section neglecting rotations and spin-correlation effects.
The YK theory shows an important deviation of the up-scattering cross section from a pure $\sqrt{T}$ dependence, which will be demonstrated experimentally in the present paper. Molecular vibrations are also included in the YK model, however, they are negligible in our experimental temperature range (vibration modes are populated only above 4000 K).

\begin{figure}
    \centering
    \includegraphics[width=1\linewidth]{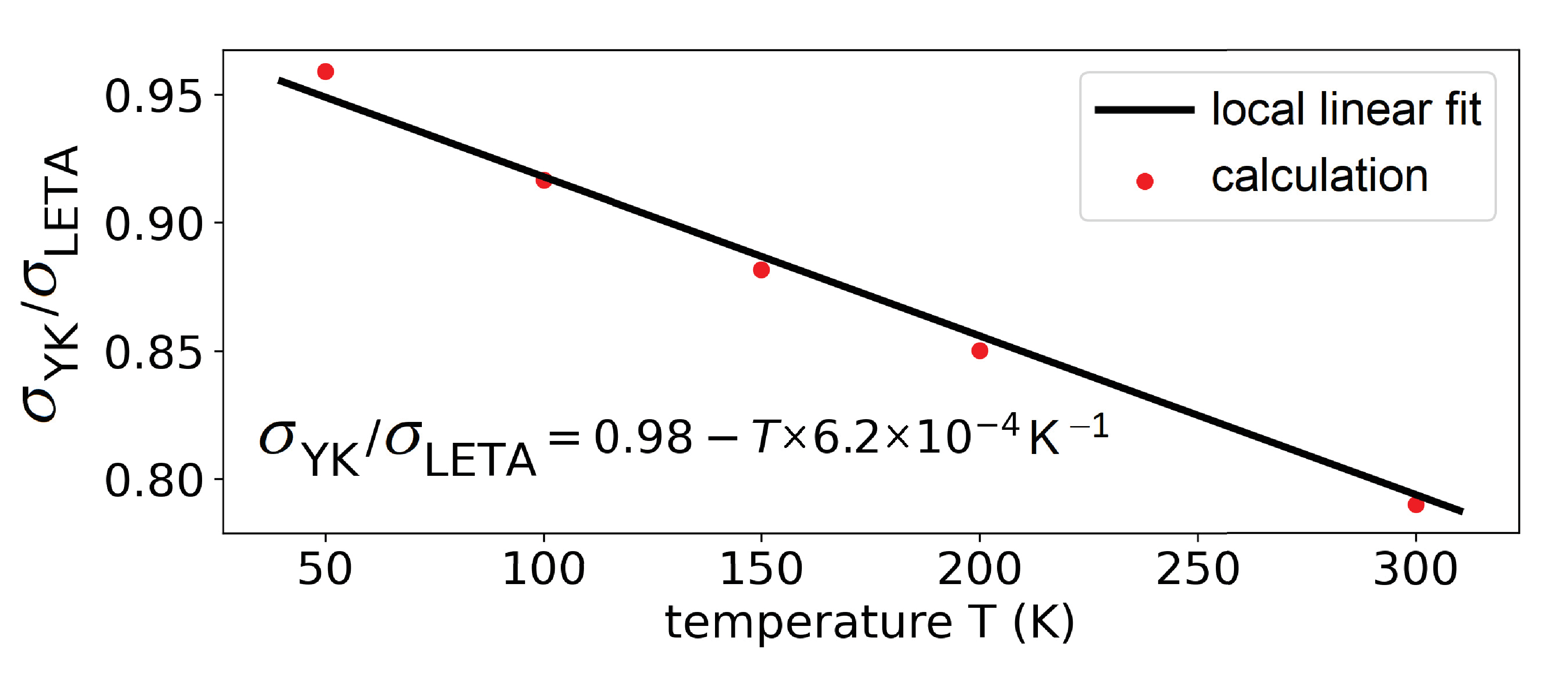}
    \caption{Ratio of the UCN upscattering cross sections calculated from the YK  and LETA models revealing the discrepancy between the temperature dependences, and a local linear fit approximation with its parameters.}
    \label{fig:Ratio-UCN-losses-upscattering-YK-and-LEA}
\end{figure}

The total upscattering cross section as a function of UCN energies can be calculated from 
Eqs.~(A4) and (A5)  in~\cite{YoungKoppel1964} by integrating over the solid angle and final energies. 
We obtained a temperature dependence characteristic of the YK model. In Fig.~\ref{fig:Ratio-UCN-losses-upscattering-YK-and-LEA} we plotted this dependence normalized with the LETA model, which only involves  $T^{1/2}$ dependence. 
At low temperatures, below the linearity range shown, the two model predictions coincide, however, in the temperature range reaching above RT there is an important deviation. The YK model cross section is about 80\% of that of LETA at RT. 

%%%%%%%%%%%%%%%%%%%%%%%%%%%%%%%%%%%%%%%%%%%%%%%%%%%%%%%%%%%%%%%%%%%%%%%%%%%%

\section{Experiment}
\label{sec:concept}
\label{sec:exp}

The experiment was designed to be sensitive to the difference 
between the monoatomic LETA and diatomic YK models,
by measuring the transmission probability of UCNs 
passing through a gas as a function of pressure and temperature.
%This fact determined our experimental and analysis strategies.

\subsection{Experimental setup}

\begin{figure}
    \centering
    \includegraphics[width=1\linewidth]{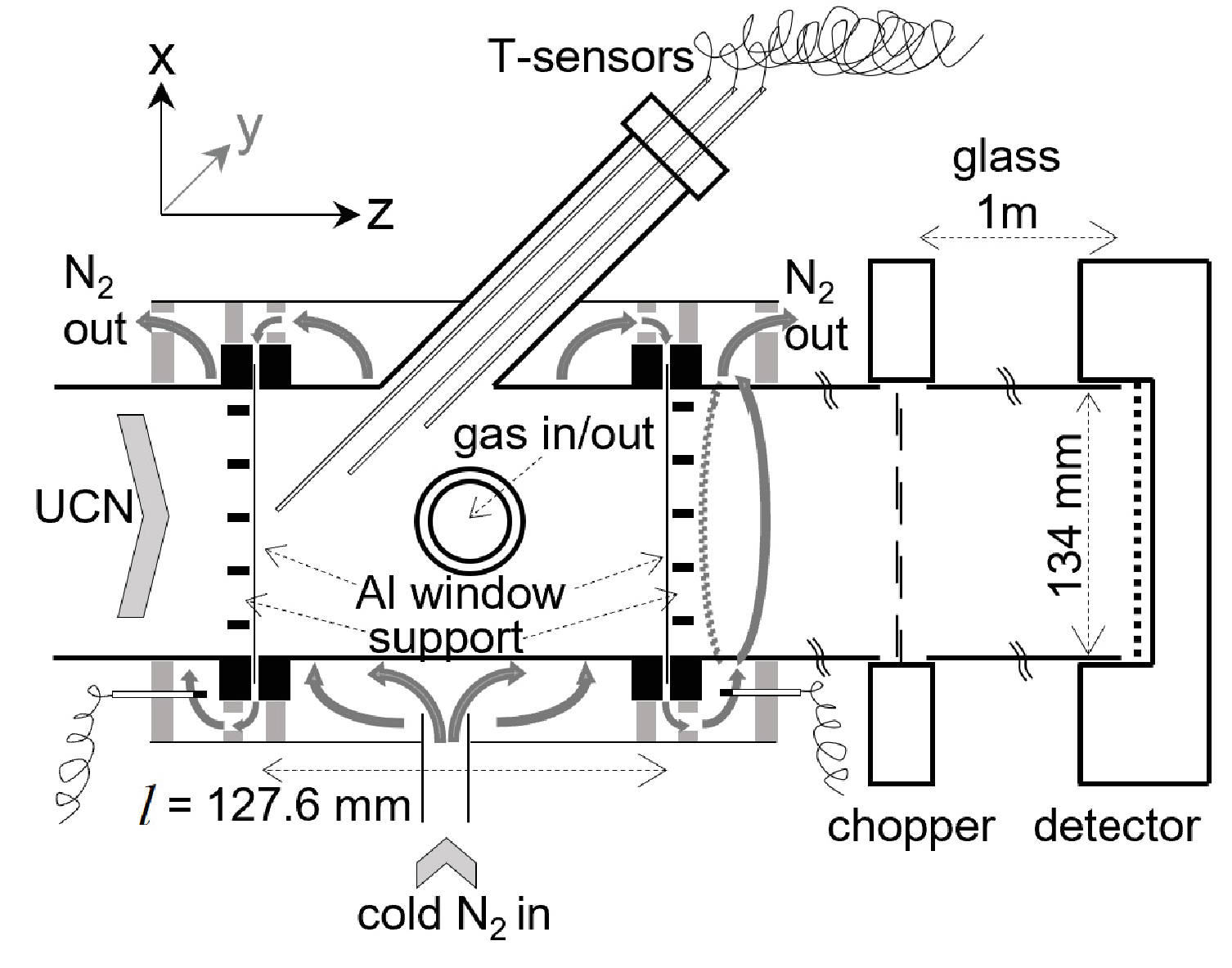}
    \caption{The %UCN-D$_2$-scattering 
		gas cell, chopper, and UCN detector (left to right, not to scale) in the standard TOF measurement. 
		On the left side, the system was connected to the beamport West-1 of the UCN source.}
    \label{fig:setup}
\end{figure}

The experiment was conducted at beamport West-1 
of the PSI UCN source, working with typically 8\,s proton pulses on the target which start UCN measurements lasting for 300\,s until the next pulse~\cite{Bison2020,Bison2022,Lauss2021scipost}.
% of the Paul Scherrer Institute~\cite{Lauss2021scipost}.
%As the main UCN-flux branch the West-1 beamport was selected and for 
Beamport South was used to monitor the UCN source stability.
We employed glass tubes coated with nickel-molybdenum (NiMo) and polished stainless steel tubes as UCN guides.
%The next parts were made from stainless steel: 
The gas cell was followed by a UCN chopper~\cite{Bison2023TOF}, 
a 1\,m UCN guide, and a CASCADE UCN detector~\cite{CASCADE2023} (see Fig.~\ref{fig:setup}). 

%\subsection{UCN-D2-scattering-cell}

%Figure~\ref{fig:setup} sketches the setup.
The gas cell was connected with the beamport to the left
and a chopper connecting via a 1\,m UCN guide to the UCN detector as in a standard time-of-flight setup. This chopper was only used for calibration measurements, otherwise left permanently open, thus ensuring stable conditions for the transmission measurements.

%On the left hand side, the UCN-D$_2$-scattering cell (see fig. \ref{fig:setup}) is connected to the West-1 beamport and on the right hand side to a linear shutter UCN (VAT~\cite{VAT2023}) before the 1~m neutron guide.
The gas cell was sealed on both sides with 0.1\,mm thick aluminum (AlMg3) windows to separate the gas volume from the vacuum in the UCN guides.
This window was supported by stainless-steel grids that covered approximately 10~\% of the tubes cross section. 
The 
%UCN guides are directly connected with the cell and the 
gas cell was made of stainless steel as a compromise between temperature stability and UCN guiding properties.
Cooling was done with a cold gas system where the cold N$_2$ is guided first around the center of the cell and then in parallel to both ends. 
To achieve a good temperature equilibration with the windows, 
it was important that the cooling gas also cooled the stainless-steel UCN guides 
connected to both sides of the cell. 
%in addition to the central cell body.
%Pre-investigations by 
Calculations and finite element simulations showed that the highest temperature gradient was expected 
starting from the center of the window %(``hot'') 
to the point where the cold N$_2$ gas is first reaching 
the cell.
Therefore, we positioned several temperature sensors inside the gas volume which were introduced 
from the top via a pipe at a 45-degree angle to the horizontal.
Additional temperature sensors were used on the outside of the cell on both aluminum windows.
The $T$-sensors were calibrated beforehand with liquid N$_2$ and distilled ice water.
In the final analysis, a time- and position-averaged temperature value of the sensor readings was used.
We also checked, that the impact of temperature gradients on the final result is negligible 
compared to our main systematic uncertainty in the UCN dwell-time spectrum.
Pressure sensors and the gas inlet were connected by a separate flange. 
%\textcolor {red} {For the pressure sensors a two point calibration was done using meteorologic reference data for atmospheric pressure and while evacuating with a two stage pump train (both with a systematic error smaller than 0.1\%).}
%\textcolor {red} {
For the pressure sensors a two point calibration was done at atmospheric pressure (using meteorologic reference data) and after evacuation with a two stage pump train (both with a systematic error smaller than 0.1\%).
%}
%
We also made sure that there was no relevant effect of a temperature gradient on the pressure measurement.

\subsection{Analysis concept}

%The experiment and data analysis was devised in a way to measure the difference between the YK and  and  model assumptions displayed in Fig.~\ref{fig:Ratio-UCN-losses-upscattering-YK-and-LEA}.

The survival probability for a single UCN with a dwell time $t$ and mean lifetime $\tau$ inside the gas, when crossing the cell, is given by:
\begin{equation}
\label{equ:probdwell}
    P = \exp(-t/\tau).
%    P_\mathrm{gas}^{t}(\tau; t) = \exp(-t/\tau)
\end{equation}
In an ideal system, the cell length, $l$, the velocity in axial direction, $v_z$, and the dwell time, $t$, are connected by the simple relation 
\begin{equation}
    t=l/v_{z}
\end{equation}
The lifetime of the UCNs inside the gas cell is given by:
\begin{equation}
    \tau = ( \rho \text{ } \sigma \text{ } v_{\mathrm{n}})^{-1}=\left(\frac{p}{k_\text{B}\sqrt{T}} \text{ } \epsilon\right)^{-1},
\end{equation}
where on the left-hand side, $\rho$ is the number density of the gas molecules,  $v_{\mathrm{n}}$ the neutron velocity and $\sigma = \sigma(T, v_{\mathrm{n}})$ the microscopic up-scattering cross section, which is inversely proportional to  $v_n$ in both YK and LETA models. On the right-hand side, we used the ideal gas law (valid at our experimental conditions), where $p$ is the pressure in the cell, $k_\text{B}$ the Boltzmann constant, and introduced the  parameter $\epsilon$  defined as:
%\begin{equation}
%    \sigma(T, v_{\mathrm{neutron}})=\frac{\sqrt T}{v_{\mathrm{neutron}}}\epsilon
%\end{equation}
\begin{equation}
\label{equ:epsdef}
    \epsilon = \frac{ \sigma \text{ } v_{ \mathrm{n} }} {\sqrt T}
\end{equation}

We introduced this scaled cross section parameter to make the analysis sensitive to temperature dependence other than $\sqrt{T}$. In case that LETA is valid, $\epsilon$ would be independent of temperature. In case the YK model is valid, $\epsilon$ would depend linearly on $T$ (in the range of interest), see Fig.~\ref{fig:Ratio-UCN-losses-upscattering-YK-and-LEA}.

Putting everything together gives a $v_z$-dependent model for the analysis of the transmission probability
\begin{equation}
\label{equ:toysingle}
    P_\mathrm{gas}^{v_{z}}(l, p, T; \frac{\epsilon}{ v_{z}}) = \exp{\left(-\frac{lp }{ k_\text{B}\sqrt{T}} \text{ } \frac{\epsilon}{v_{z}}\right)},
\end{equation}
where the first three arguments describe the experimental conditions for the gas.

The quantity $\epsilon$ is independent of temperature for monoatomic gases like Ne and is temperature dependent for diatomic gases like D$_2$.
As indicated in Fig.~\ref{fig:Ratio-UCN-losses-upscattering-YK-and-LEA} for D$_2$, the ratio of cross sections calculated with the YK and LETA models (thus also the ratio of the $\epsilon$-s defined via Eq.~\ref{equ:epsdef}) can be approximated with a linear $T$-dependent fit function in the temperature range of interest (expecting discernible effects):
\begin{equation}
\label{equ:et}
    \epsilon(T)=\epsilon_0 \text{ } (\epsilon_a - \epsilon_T \text{ } T),
%\label{equ:epsilon_vs_T}
\end{equation}
where $\epsilon_0$ represents the LETA approximation, 
and $\epsilon_a =0.98\pm 0.03$ and $\epsilon_T = (6.26\pm 0.37) \times 10^{-4} $ K$^{-1} $. 
%if the YK model is valid, and $\epsilon_T=0$ if LETA is valid.
Besides the fit error, there is a $\sim 1~\%$ error from the numerical uncertainty of the MC integration and $\sim 2~\%$ from uncertainty in the input parameters like the equilibrium distance of the nuclei.
In total, we estimate the slope error to be below $6~\%$.
%An accurate measurement of the slope and absolute cross section could also help improve the theoretical accuracy.

Other than in the theory, in the analysis of the  measurements, one cannot disentangle $\epsilon_a$ from $\epsilon_0$. 
However, we were looking for the temperature dependence additional to $\sqrt{T}$.
%\textcolor{red}{
The signature we were searching for was covered by $\epsilon_T$. We thus set $\epsilon_a$ to the theoretically expected value of $0.98$ valid in our temperature range of interest, and propagate its uncertainty into the final systematic error.
%}

In our analysis, we used calibrated (see next section) values for $l$, $p$ (0 to 1 bar), $T$ (range for all gases: 110~K to 365~K) and integrated Eq.~\ref{equ:toysingle} over the normalized neutron velocity distribution $f(v_z)$ which was extracted from the combined time-of-flight (TOF) measurements and D$_2$ calibration (see Fig. 3 and details below) to obtain a measurable transmission probability
\begin{equation}
    P_\mathrm{gas}(l, p, T; \epsilon) = \int P_\mathrm{gas}^{v_{z}}(l, p, T; \frac{\epsilon}{ v_{z}}) \text{ }  f(v_z) \mathrm d{v_z},
\label{equ:pgas}
\end{equation}
where $\epsilon$ also involves the parameter $\epsilon_\text{T}$.
As explained later, to also consider unavoidable UCN storage effects in the cell, integration over the velocity distribution was replaced in the analysis by the equivalent integration over dwell time (see Eq.~\ref{equ:probdwell}).%distribution by utilizing the relation $N(v_Z) \mathrm d{v_Z}  = N'(t_\mathrm{dwell}) \mathrm d{t_{dwell}} $.

We extensively investigated the measured data sets of two gases: neon (Ne) to serve as a test candidate for the LETA model and D$_2$ for the YK model.
We checked whether $\epsilon_T$ is compatible with $0$ in the case of Ne,  and whether for D$_2$ it is compatible with the value $(6.26\pm0.37) \times 10^{-4}\,\mathrm{K}^{-1}$ from the theoretical Young-Koppel prediction.

As an additional check of systematics, mainly of the dwell time spectrum, we also measured He, Ar, CF$_4$, Xe and H$_2$ gases, but in a reduced $p-T$ parameter space.

% (((COMMENT from Geza: in this list below, that I commented out,  the UCN-related part is understandable only if we first show the experiment setup, and the rest was already mentioned above. What was not mentioned, I inserted above. So we don't repeat things in a concise paper)))
%The needs for the experimental setup were therefore:
%\begin{enumerate}
%    \item UCN with the measurement of: 
%    \begin{enumerate}
%       \item count-rates in main detector 
%        \item monitoring detector for source stability
%        \item dwell time spectrum obtained by time of flight measurements
%    \end{enumerate}
%    \item scattering cell for gaseous samples:
%    \begin{enumerate}
%        \item species: D$_2$, Ne
%        \item H$_2$, Xe... for additional systematic crosscheck
%        \item pressure: 0 to 1 bar
%        \item temperature: 100 K to 400 K
%    \end{enumerate}
%    \item and pressure and temperature measurement.
%\end{enumerate}

%----------------------------------------------------------------------------------------------
\section{Systematics}

%\subsection{Measurement of $P_{\mathrm{gas}}$ }

To extract the transmission probability function $P_{\mathrm{gas}}$ from the measurements, four major systematic effects had to be taken into account:
\begin{itemize}
    \item time variation of the UCN rate at the beamport~\cite{Anghel2018},
    \item faster neutrons during each beam pulse~\cite{Bison2020},
    \item temperature-dependent transmission probability of the empty cell, mainly of the windows,
    \item dwell time of UCNs in the gas cell.
\end{itemize}

A second CASCADE UCN detector (labeled here "B")
on beamport South was used
to monitor the UCN rate (measured with detector "A") at the West-1 beamport.
% as a monitor for the total amount of neutrons produced.
Cutting the first 12\,s after the start of the proton beam pulse
eliminated the contribution of neutrons faster than UCN~\cite{Bison2023TOF}.
%The neutrons during the proton beam pulse 
%were rejected by cutting the UCN data of the first 12~\,s after each proton pulse~\cite{Bison2023TOF}.
%
In this way the ratio  
$R$ 
between the neutron count rates of the two detectors for the empty cell,
%at room temperature 
%where $T_{ref}$-s are a set of temperatures measured with an empty cell, 
was determined to be constant at the $<1\,\%$ level.
%
%The fluctuation in time is dominated by the count rate statistics of the two detectors.
% and has been shown in previous publications -- not found where published, rather not published (Geza). [ZITATE]

The temperature-dependent transmission behavior of the empty cell was taken care of by calibration, as described in the second subsection below, applying a temperature-dependent correction factor for the total transmission probability.

Taking these three systematic effects into account $P_{\mathrm{gas}}(T)$ can be written as:
\begin{equation}
    P_{\mathrm{gas}}(T)= \frac{1}%{R(T_\mathrm{ref}) \cdot  
    {R \text{ }  P_\mathrm{window}(T)} \text{ }  \frac{N_A(p,T)}{ N_B}, 
\end{equation}
where $N_A$ is the neutron count of the main detector and $N_B$ is the neutron count of the monitoring detector for each proton beam pulse (both without the first $12\,\mathrm{s}$).
$P_\mathrm{window}(T)$ represents the temperature-dependent transmission probability 
of the empty cell and the UCN guides.
The reference value 
%$R(T_\text{ref})$ 
$R$ 
includes all other constant effects that influence the ratio between the number of counts in the detectors A and B.

%---------------------------------

%\section{Systematics}

\subsection{Reference value of the setup with empty cell}

The reference 
%$R(T_\mathrm{ref})$ 
$R$ 
was measured 
periodically
%for every campaign at least twice, 
after and before thermal annealing of the D$_2$ crystal~\cite{Anghel2018,RienaeckerThesis2022} used for UCN production,  for several UCN pulses with an empty cell. 
As the $T$-dependence of the UCN guides and cell without test-gas was reproducible, it could be done 
at any temperature $T_\text{ref}$
of the empty gas cell.
%Concerning the temperatures of the empty cell measurements, $T_\text{ref}$, 
%
%
However, most of the time it was measured at room temperature.
The probability  $P_{\mathrm{gas}}(T)$ with an empty cell must be per definition 1, therefore, %$R(T_\mathrm{ref})$
$R$
can be written as:
\begin{equation}
%    R(T_\mathrm{ref})=\frac{N_{A,\mathrm{ref}}(p=0,T_\mathrm{ref})}{P_\mathrm{window}(T_\mathrm{ref})} \cdot \frac{1}{N_{B,\mathrm{ref}}}
    R=\frac{N_{A,\mathrm{ref}}(p=0,T_\mathrm{ref})}{P_\mathrm{window}(T_\mathrm{ref})} \text{ } \frac{1}{N_{B,\mathrm{ref}}}
\end{equation}

\subsection{Temperature dependence for the empty cell}

For the empty cell 
%and main beam line 
we found a thermal correction function by fitting the transmission data:
\begin{equation}
    P_\mathrm{window}(T) = \frac{C_T \text{ } T +C_0}{C_0},
\end{equation}
%(((Was ist $C_0$? Warum brauchen wir oder die Leser 3 Parameter?))) --> Tippfehler beides das selbe C0
where $C_T = (-1.73\pm 0.03) \times 10^{-3} K^{-1} $ and $C_0 = 5.470\pm 0.006$. %(((Gross C oder klein c?))).
Since it is impossible to disentangle the absolute transmission probability of the setup, the function was built in a way that it is $1$ at $0\,\mathrm{K}$.
%(((Geza versteht diesen Satzteil nicht:))) 
%
This is valid since the absolute scale will cancel out in $P_\mathrm{gas}$ with the absolute scale of $R(T_\mathrm{ref})$.
The relevant part for this investigation is, 
that the empty cell transmission changes with the temperature 
so that the transmission probability of the empty cell at around $100\, \mathrm K$ is about $6~\%$ 
higher than at RT. This is plausible considering up-scattering on phonons in the 
aluminum window
causing a previously observed temperature dependence of the total UCN cross section 
on aluminum~\cite{Steyerl1971,Steyerl1972b}. 
%Das wäre noch eine Frage an Ingo, ich denke er hatte das damals gerechnet.

%RG: Hoffe der Abschnitt ist auf diese Art klarer erklärt. Passt das so für Dich? @Geza

\subsection{Pressure-induced change of cell length}

The effective length of the gas cell changes with increasing pressure
in the cell,
due to a small bulging of the aluminum foils on the grid. 
The size of this effect was estimated via a measurement of
the pressure change in the gas cell while on the other 
sides of the windows the 
pressure changed from 1\,bar to 0 during pumping. 
%
%We did measure the average effect by a dedicated setup where we used the cell, and closed the ends.
%By this, we could form three chambers, with the main scatter cell in the center, separated by the aluminum foil windows.
%During the whole time, pressure in all three compartments was monitored and corrected for temperature fluctuations.
%The outer two compartments (oc) were connected and then slowly pumped down inducing a pressure difference on the aluminum windows.
%During this, we could observe a small pressure drop in the central chamber (cc) which we interpreted as the bending of aluminum windows.
At a starting pressure of $972.0\,\mathrm{mbar}$ a small but well-measurable drop to $964.5\,\mathrm{mbar}$ was observed and could be reproduced several times.
This was interpreted by ideal gas law as a linear change in the average cell length of 0.8~\% 
from 0 to 1\,bar with a linear dependence on the pressure.
%$10^5\,Pa$. 
%This effect would be negligible for us.
%However, we found a very good reprehensibility of this effect and it was in agreement with optical inspection of the bending of the windows so it was included in the analysis.

Therefore, the effective mean length of the cell can be written as a function of pressure:
\begin{equation}
\label{equ:lengthVSpressure}l = 127.6\, \mathrm{mm} + p \times 1.060~\mathrm{mm/bar}    
\end{equation}

This was in agreement with optical inspection and due to its small size a minor contribution
in our analysis.

\subsection{Calibration of the dwell-time spectrum} %, time of flight and axial velocity, $v_z$}

\begin{figure}
    \centering
    \includegraphics[width=1\linewidth]{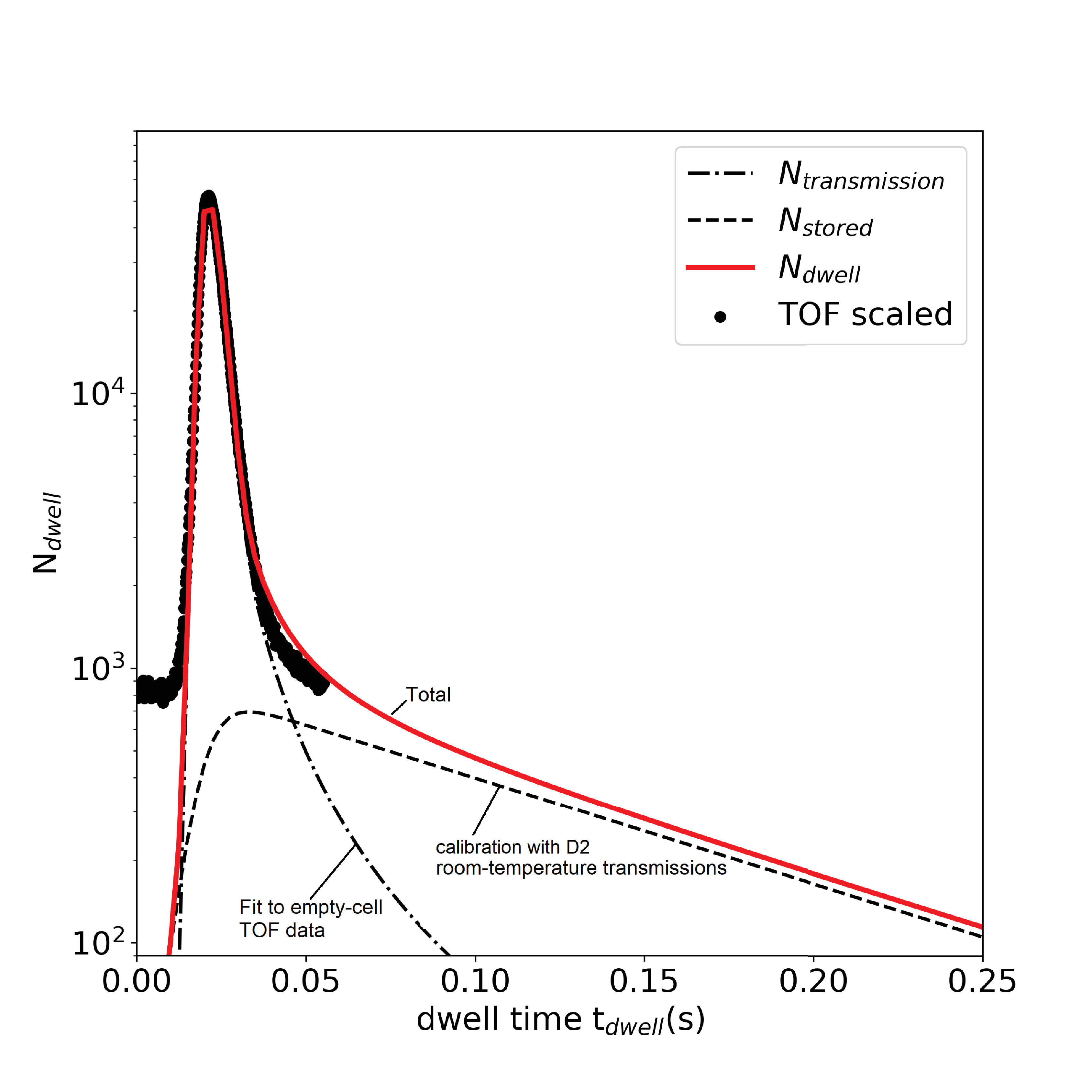}
    \caption{ 
    The dwell time spectrum of UCNs 
    %in the analysis time window t$>$12\,s after the start of the proton pulse 
    is composed of two parts. 
    The time axis was scaled using the ratio of the effective length of the gas cell and the flight path.
    %that we derived independently.
    First part: The time-of-flight spectrum of raw neutron counts measured with an empty cell (black dots `TOF scaled'). 
    %The y-axis can be understood as the raw neutron counts per bin (no scaling). 
    The fit is denoted by `$N_\text{transmission}$'.
    %Absolute values will be irrelevant later on since they disappear through normalizations, however, they are an indicator for statistical error estimation.
    Second part: 
    %In our setup we were blind to stored neutrons in the cell since the chopper, flight path and detector are behind the cell.
    %Therefore we had to indirectly obtain the 
    contribution of stored neutrons obtained from calibration to the RT data of D$_2$ denoted as `$N_\text{stored}$'.
    In the analysis 
    %of our transmission data 
    we used the combined spectrum `$N_\text{dwell}$' (see text).
    }
    \label{fig:dwelltime}
\end{figure}

%In addition to the three systematic effects that can be easily taken into account, there is a fourth systematic effect that we found by calculation and simulation to be a major systematic contribution.
%It is the knowledge of the time each neutron stays inside the scattering cell the so-called dwell time or?? needed for the analysis is the dwell time distribution $N_{\mathrm{dwell}}$.

In our analysis the knowledge of the dwell-time distribution, $N(t_\mathrm{dwell})$, i.e. the time UCNs spend inside the gas cell, is needed.
In an ideal setup, where the axial velocity of the neutrons i.e. in the direction of the UCN guide axis,
$v_{z}$, does not change over time, $t_\mathrm{dwell}$ can be derived from $v_{z}$: 
\begin{equation}
    t_\mathrm{dwell} = \frac{l_\mathrm{cell}}{v_z},
\end{equation}
or $v_{z}$ can be measured via time-of-flight (TOF) from the chopper to the detector:
\begin{equation}
    v_{z} = \frac{l_\mathrm{TOF}}{t_\mathrm{TOF}}.
\end{equation}

%o adding in a time of flight measurement beyond the cell allows for deriving $N_{\mathrm{dwell}}$ in an ideal setup.

%n our setup, we measured the time of flight distribution with a chopper and detector behind the scattering cell.
The UCN time-of-flight distribution was first measured with the 
chopper
installed after the gas cell as in a standard TOF measurement (Fig. \ref{fig:dwelltime}).
% and
%and the velocity distribution of the UCNs was derived.
%hus we could get a good understanding of the velocity distribution of the UCNs.
However, this setup is blind to neutrons that are stored for a considerable time within the gas cell due to diffuse reflections, 
%(path~$> l_\text{cell}$)
which causes an additional systematic. 
% so we needed to obtain more information.

%--------------------------------------

%There were several pieces of evidence for a storage effect like a very specific signature in the residuals of all fits, the cross-section for D$_2$ and Ne did depend strongly on the selection of a subset of the data, the cross-sections of all test gases where differing to literature by a range of 10\% up to a factor of two.
%With this evidence, we decided to have a test experiment dedicated to measuring the storage effect of our scattering cell. 
%Therefore the setup was rearranged:
%1. UCN source, 2. chopper, 3. scatter cell and 4. detector A.
%With this setup, we could prove that there is an exponential tail to the UCN spectrum that we have been blind to when utilizing the former setup.

%\textcolor{red}{OLD: To determine the size of this effect, the gas cell was placed between the chopper and the detector.
%The observed time-of-flight distribution showed an exponential tail in the UCN spectrum that we interpreted as being due to longer stored UCNs inside the cell.
%we have been blind to when utilizing the former setup. Figure~\ref{fig:dwelltime}) shows the extracted dwell-time distribution described by a modified asymmetric Gaussian with the main peak around $~25\,\mathrm{ms}$,  which corresponds to a neutron velocity of $5.2\,\mathrm{m/s} $. We checked for the impact of several factors like the precise shape of the main peak or small deviations at the rising and falling flanks,  and found those to be negligible in our case.}

To determine the size of this effect, we made use of the RT subset of all D$_2$ transmission data measured with the chopper permanently open. 
First, from the TOF data taken with an empty cell, we performed an analytical parametrization of the main peak of the UCN spectrum. The profile can be described by a modified asymmetric Gaussian with a main peak around $~0.025\,\mathrm{s}$. This corresponds to a neutron velocity of $5.2\,\mathrm{m/s}$ 
(see TOF fit curve `$N_\text{transmission}$' in Fig.~\ref{fig:dwelltime}).
We checked for the impact of several factors like the precise shape of the main peak
or small deviations at the rising and falling flanks, 
and found those to be negligible in our case.

As the second step, to describe the part of stored neutrons that are not visible in the TOF data we added an exponential tail:
%\textcolor{red}{OLD: 
%To describe the stored neutrons we added as an additional component an exponential decay:}
\begin{equation}
    N_\mathrm{stored}= a_{storage} \text{ } \exp{(-t_\mathrm{dwell}/\tau_{storage})}
\end{equation}
and on the left side a rising edge of the stored UCN curve which
%beginning of this part 
is described with a logistic growth function to emulate the filling of the system. 
The impact of the precise shape of the latter was found to be negligible. 
%Absolute count values will be irrelevant later on since they disappear through normalizations, however, they are an indicator for statistical error estimation.
The final curve for the stored UCNs with calibrated parameters is shown as `$N_\text{stored}$' in Fig.~\ref{fig:dwelltime}).  
The calibration of the exponential constants was determined by a global fit of Eq.~\ref{equ:pgas} to all room-temperature D$_2$-transmission data:
%by variation of $a_{storage}$ and $t_{storage}$:

%\begin{equation}
%    t_{storage}= 112.5\pm 8.0 ~\mathrm{ms},  \\ 
%		a_{storage}= 968\pm 115 
%    \label{equ:stored}
%\end{equation}

%\begin{equation}
\begin{align}
\begin{split}
    \tau_{storage}= 112.5\pm 8.0 ~\mathrm{ms},  \\ 
		a_{storage}= 968\pm 115. 
    \label{equ:stored}
\end{split}
\end{align}
%\end{equation}
Absolute amplitude values cancel out through later normalization, however, they are an indicator for statistical error estimation.

%----------------------------------
By this, we have a complete and calibrated description of the dwell-time spectrum.
This was used in the analyses of all investigated gases at all temperatures and with both models (LETA and YK).
%For sure, for a new experiment, the dwell time spectrum would be of major interest to improve systematics and absolute cross sections and a future improved setup should be designed accounting for this.

As an independent check of this exponential part, we rearranged the setup so that the chopper was placed before the cell.
With this setup, we found a storage time constant of UCNs with around $92\,\mathrm{ms}$, which is 
%within 2.6$\sigma$ and 
in reasonable agreement with the time constant, we found through the calibration procedure with D$_2$ gas. 

\subsection{Systematic test with reference gases}

To check our understanding of the dwell-time, we
measured UCN transmission of gases as He, Ne, Ar, CF$_4$, Xe as references,
and derived the
total neutron cross section for a given gas at a given temperature.

%The effect of the neutron dwell time can be understood in a way that UCNs with long dwell times are sensitive to long lifetime constants and short dwell times to short lifetime constants.
%In our case this resulted in several observations when neglecting the stored UCNs (long dwell times):
%First, the total cross-section for a given gas at a given temperature was pressure-dependent. 
%Second, when using different gases with different cross-sections the relative deviation between literature and measurement depends on the cross-section.
%For some gases, we found deviations of more than a factor of two for others the deviation was some percent compared to literature values.
%Third, when looking at the residuals of each exponential fit there was a very characteristic signature that could be reproduced by simulating the transmission curve with a stored neutron tail but neglecting it during the fit procedure.

Again, accounting for the correct dwell-time spectrum including stored UCNs we found good agreement with the literature and calculated the values shown in Table~\ref{tab:screening}. 
%\textcolor {red} {
We also obtained very good agreement with the measurements of Ne, Ar, Xe, CF4 in~\cite{SeestromEtAl2015upscattering} at RT, the difference being smaller than the error bars.
%}
%\textcolor{blue}{
For reference, we considered 6.6\,m/s UCN-velocity in order to compare to previous 
work~\cite{SeestromEtAl2015upscattering} applying the same textbook equation
(Eq.~(11) in \cite{SeestromEtAl2015upscattering} from Ref.~\cite{Turchin1965}) 
to convert from free atoms at rest to a Maxwell-Boltzmann gas.
%}
The values for the total cross section were obtained by fitting Eq.~\ref{equ:pgas} 
to transmission data using the dwell-time spectrum from Fig.~\ref{fig:dwelltime}.
The statistical uncertainty is extracted from the fit.
The systematic uncertainty was calculated using maximum error propagation by fitting with 
minimum and maximum values derived from  
the uncertainties 
on $\tau_{\mathrm{storage}}$ and $a_{\mathrm{storage}}$ (Eq.~\ref{equ:stored}). 
%and refitting with those.
%The difference 
%%between maximum and minimum fit results 
%is %then 
%interpreted as the systematic confidence interval.

By this, we find that for all investigated gases 
the systematic uncertainties dominate the statistical ones by at least one order of magnitude.
%
%For a new measurement campaign, it would be therefore crucial to improve the knowledge of the dwell-time spectrum to reduce the resulting error on absolute cross-sections.

%\begin{table}[htbp]
%  \centering
%  \caption{...The literature values for upscattering with exception of CF$_4$(\cite{SeestromEtAl2015upscattering}) were calculated from tabulated data~\cite{VarleySears1992neutronXS} and with the YK model of H$_2$ and D$_2$  for 6.6 m/s and at T=294.5 K. Absorption data were all calculated from~\cite{VarleySears1992neutronXS}.}
%    \begin{tabular}{lccccc}
%    \hline
%     & \multicolumn{3}{c}{this work} & \multicolumn{2}{c}{literature} \\
%    %\cline{2-4} 
%    %\cline{5-6}
%    \cmidrule(lr{.0em}){2-4}
%    \cmidrule(lr{.0em}){5-6}
%   species& {$\epsilon$} & {stat.} & {syst.} & {$\epsilon$} & {$\sigma$} \\
%          & {($  \frac{\mu mol K^{1/2}}{m}$)} & $\pm$      & $\pm$      & {($  \frac{\mu mol K^{1/2}}{m}$)}& {$(\mathrm{barn})$} \\
%   \hline
%    He    & 4235  & 26    & 390   & 3763  & 164 \\
%    Ne    & 5856  & 13    & 513   & 4911  & 214 \\
%    Ar    & 6596  & 40    & 571   & 6058  & 264 \\
%    D$_2$    & 62434 & 100   & 2703  & 61956 & 2700 \\
%    CF$_4$   & 79241 & 43    & 3242  & 75725 & 3300 \\
%    Xe    & 188733 & 783   & 8922  & 185020 & 8063 \\
%    H$_2$    & 399353 & 1141  & 22225 & 367149 & 16000 \\
%    H$_2$    & 399353 & 1141  & 22225 & 414419 & 18060 \\
%    \hline
%    \end{tabular}%
%  \label{tab:screening}%
%\end{table}%

\begin{table}[htbp]
  \centering
  \caption{Measured total cross sections calculated for 6.6~m/s and at T=294.5~K including statistical and systematic uncertainties. Most `literature' values were calculated from tabulated data~\cite{VarleySears1992neutronXS}. 
 Absorption data were all calculated from Ref.~\cite{VarleySears1992neutronXS}. The CF$_4$ cross section is cited from (\cite{SeestromEtAl2015upscattering}).
The `literature' cross sections for D$_2$ and H$_2$ were calculated in this work by applying the YK model. 
%\textcolor {red} {CORRECTED D$_2$ and Ne "syst."}
  }
    \begin{ruledtabular}
    \begin{tabular}{lccccc}
%    \hline
     & \multicolumn{3}{c}{this work} & \multicolumn{2}{c}{literature} \\
    %\cline{2-4} 
    %\cline{5-6}
    \cmidrule(lr{.0em}){2-4}
    \cmidrule(lr{.0em}){5-6}
   species& {$\sigma$} & {stat.} & {syst.} &  {$\sigma$} & \\
          &  barn & $\pm$ & $\pm$ & barn &  \\
   \hline
He & 185 & 1.13 & 17.0 & 164 \\
%Ne & 255 &	0.566 &	22.4 & 214   \\
Ne & 255 &	0.566 &	23.7 & 214   \\
Ar & 287 &	1.74 & 24.9 & 264   \\
Xe & 8220 &	34.1 & 389 & 8063   \\
CF$_4$ & 3450 &	1.87 & 141 & 3300   \\
%D$_2$ & 2626 & 18 & 122 &	2582   \\
%D$_2$ & 2626 & 97 & 157 &	2582   \\
D$_2$ & 2626 & 18 & 157 &	2582   \\
H$_2$ & 17400 & 49.7 &	969	& 18060   \\
%    \hline
    \end{tabular}%
    \end{ruledtabular}
\label{tab:screening}%
\end{table}%

%---------------------------------------------------

\section{Results and discussion}

The main transmission measurements were performed with gaseous D$_2$ 
%\textcolor {red} {
(99.6~\% D$_2$ purity with a fraction of HD molecules of 0.4~\% which has a negligible contribution compared to the main systematic effects discussed)
%}
 and Ne in the pressure 
range from 0 to 1\,bar, 
%regime from 0 to $101\,\mathrm{k Pa}$ 
and from $110\,\mathrm{K}$ to $365\,\mathrm{K}$.
%
%\textcolor {red} {es braucht einen Satz zur x Achse in den PLots}
%\textcolor{blue}{
The pressure, $p$, and the temperature, $T$, of the gas are measured directly with sensors in each measurement.

\begin{figure}
    \centering
    \includegraphics[width=1\linewidth]{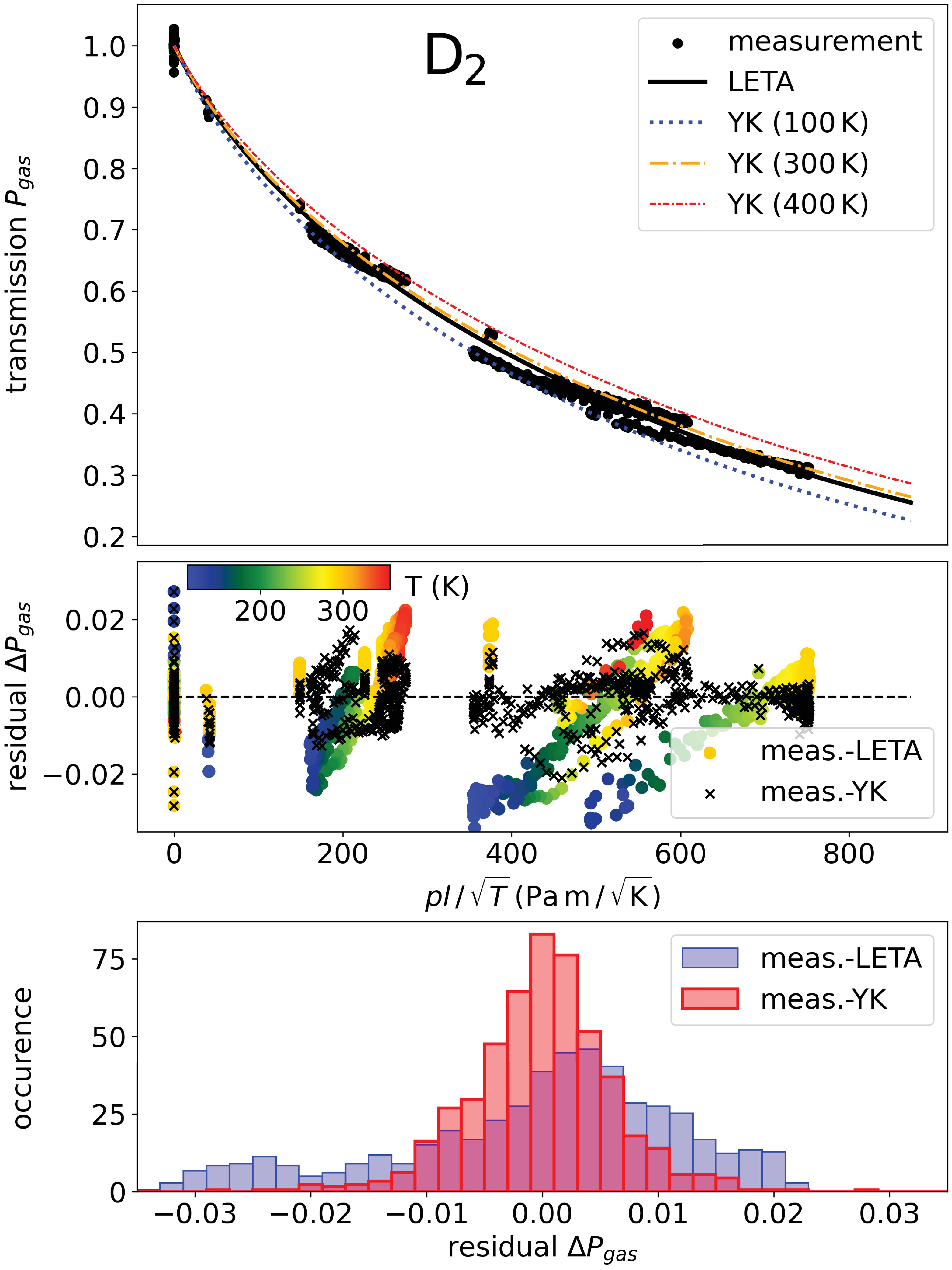}
    \caption{Top: UCN transmission data (black dots) in D$_2$ gas as a function of ${\left(\frac{lp }{\sqrt{T}}\right)}$ (at all measured $T$-s), and
	the best fits of the two alternative models with example values for the independent variable $T$ in case of YK (blue=100~K, yellow=300~K, red=400~K lines), and one line with LETA (black). 
	Center: Point-by-point residuals of the YK and LETA model fits to the measured data, indicating smaller deviations for the YK model fit, showing data at all measured temperatures in a color scale.
	Bottom: Histograms of the above residuals showing a much narrower and symmetrical distribution for the case of the YK model.}
    \label{fig:D2trans}
\end{figure}

\begin{figure}
    \centering
    \includegraphics[width=1\linewidth]{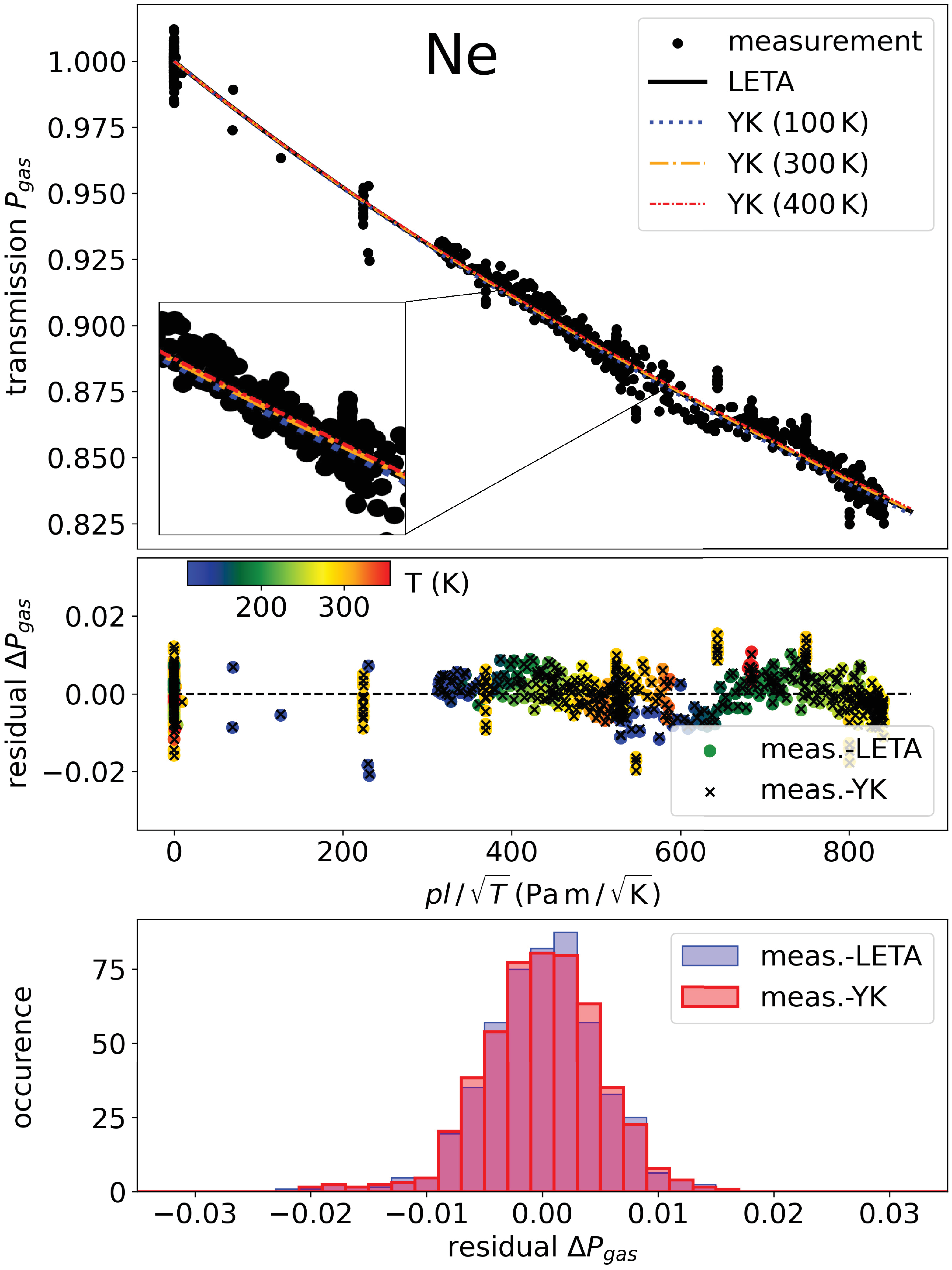}
    \caption{Top: UCN transmission data (black dots) in Ne gas as a function of ${\left(\frac{lp }{\sqrt{T}}\right)}$ (at all measured $T$-s), and
	the best fits of the two alternative models with example values for the independent variable $T$ in the case of YK (blue=100~K, yellow=300~K, red=400~K lines), and one line with LETA (black). 
		The black and colored curves overlap (see insert), since no 
		$T$-dependence of $\epsilon$ was obtained from the Ne gas fit.
		Center: Residuals of the LETA and YK model fits to the measured data, here displaying all the measured temperatures in color scale. 
		Bottom: Histograms of the above residuals.}
    \label{fig:Netrans}
\end{figure}

In Eq.~\ref{equ:toysingle} the exponent of the transmission probability of a neutron is given as 
${\left(-\frac{lp }{ k_\text{B}\sqrt{T}} \text{ } \frac{\epsilon}{v_{z}}\right)}$, 
where the cell length, $l$, is dependent on the absolute pressure, $p$, 
according to Eq.~\ref{equ:lengthVSpressure}.
Consequently, the transmission can be plotted 
as a function of ${\left(\frac{lp }{\sqrt{T}}\right)}$. 
By this choice, one expects a close-to-exponential dependence of the measured transmissions 
on ${\left(\frac{lp }{\sqrt{T}}\right)}$, which also includes the broadening due to the dwell time spectrum. These transmission profiles are seen in the top plots of Fig.~\ref{fig:D2trans} 
for D$_2$ gas and Fig.~\ref{fig:Netrans} for Ne. 

The aforementioned spectral broadening would make calculating $\epsilon(T)$ directly as a function of transmission (by inverting Eq.~\ref{equ:pgas}) to an arduous computational challenge.
Thus we developed a feasible numerical fit alternative.
In the data fit, we chose the transmission probability as the dependent variable, and the two independent variables were thus $T$ and ${\left(\frac{lp }{\sqrt{T}}\right)}$, which replaced $p$. The unknown quantity given by the two-variable fit thus was $\epsilon$.
%
%The neutron velocity distribution, $f(v_{z})$ in Eq.~\ref{equ:pgas} is extracted from our calibration measurements, converted to dwell time spectrum (see Fig.~\ref{fig:dwelltime}).
%\textcolor{orange}{
Therefore, for the LETA model a single-parameter ($\epsilon$) fit and for the YK model a 
two-parameter ($\epsilon_0$, $\epsilon_\mathrm{T}$) fit was performed. 
The latter additional parameter $\epsilon_\mathrm{T}$ gave a small modulation to the exponential dependence, due to the additional $T$-dependent term, which served as the signature of the YK model. %}
%\textcolor{violet}{
Our least-square fit minimized the sum of the transmission residuals from all measurements for a given gas at different $T$ and ${\left(\frac{lp }{\sqrt{T}}\right)}$ values. The residuals were defined here as the difference between the measured 
UCN transmission and the one calculated via Eq.~\ref{equ:pgas}, taking into account that $\epsilon$ 
has the form as in Eq.~\ref{equ:et}.

In the top of Figs.~\ref{fig:D2trans} and~\ref{fig:Netrans}, along with the measured 
transmissions, we also plotted the best fits with the two models for $\epsilon$ plugged into Eq.~\ref{equ:pgas} (corresponding to YK and LETA):  
one with $\epsilon = \epsilon(T)$ 
(Eq.~\ref{equ:et}) and 
one with $\epsilon$ independent of temperature. 
%For clarity, in the case of the YK model we plotted example fits for only 
%three temperatures 100~K, 300~K and 400~K.}
Since a full 3D plot of this two-variable fit problem would be much more difficult to interpret, we visualized the theoretical fit in the plots using only three example values for the independent variable $T$ (100~K, 300~K and 400~K, not to be confounded with measured values).
%}

%%\textcolor{orange}{
In the case of the LETA model, the horizontal axis variable ${\left(\frac{lp }{\sqrt{T}}\right)}$ is completely describing the $\sqrt T$ dependence. 
In the case of the YK model, an additional temperature dependence
occurs via the $T$-dependence of $\epsilon$. 
%%}
%\textcolor{blue}{
Thus for different $T$-s, we obtain different 
slopes for the exponent 
in the transmission versus ${\left(\frac{lp }{\sqrt{T}}\right)}$ corresponding 
to each $\epsilon(T)$.
%}
%due to the additional $T$ dependence, the main axis alone is not complete.
This results in splitting the exponential transmission curve into separate 
curves corresponding to the number of plotted temperatures. 
Exactly this splitting is the signal to distinguish between the LETA and YK models, as visualized 
in Fig.~\ref{fig:D2trans} for D$_2$ and in Fig.~\ref{fig:Netrans} for Ne. Here we plotted three lines (i.e. three 3D-slices) for YK corresponding to three example values of the independent variable $T$ enclosing the range of the measured temperatures.
% Steht schon oben
%The residuals in the second row and the histograms in the third row are calculated point by point from the measured transmission value for a $p$ and $T$, the calibrated $l(p)$, and the theoretical fit value as a function of  $p\cdot l/\sqrt(T)$.
%%}
%In addition, the corresponding residuals and their histograms are shown.

%The measured D2 are plotted in Fig.~\ref{fig:D2trans}, and the measured Ne data in Fig.~\ref{fig:Netrans}.

%In the case of the Ne data the shape of transmission probability, their corresponding residuals, and histograms are almost identical for each model (YK and LETA) used to fit the data.

%In the case of  D$_2$ there is an important difference in the shape, the residuals, and histograms between the two fit models.
%Especially for the YK fit one can see the additional T dependence in the splitting of the red lines.
%Those lines correspond to the temperatures $100\,\mathrm{K}$,  $300\,\mathrm{K}$ and $400\,\mathrm{K}$

%%In Fig.~\ref{fig:D2trans} 
%%the top plot displays one line for the LETA model (blue) and 
%%three distinct lines for the YK model (red).
%which correspond to temperatures of 
%100\,K, 300\,K and 400\,K,
%$100\,\mathrm{K}$, $300\,\mathrm{K}$ and $400\,\mathrm{K}$,
%that exemplify the range of our data.
%
% Wiederholung
%This test with fits at different temperatures gives a signature for the YK model since our choice of the horizontal axis variable only reflects the combined $p$ and $\sqrt{T}$ dependence of the LETA model and does not cover the additional linear $T$ dependence of the YK-model.
%
In contrast to D$_2$, the Ne data fits with different temperatures 
shown in Fig.~\ref{fig:Netrans} converge with a slope in $T$ 
compatible with zero, see also Table~\ref{tab:ned2comp}, 
hence the fit involving Eq.~\ref{equ:et} reproduces the LETA model.
%
%In the case of the D$_2$ data, one LETA (blue) and three distinct YK (red) lines are visible revealing the additional temperature dependence beyond LETA.

The best fit of the Ne data results 
in $\epsilon_T=(54\pm38\pm10) \times \,\mathrm{10^{-6}~K^{-1}}$, 
which is
%within $1.5\, \sigma$ 
about compatible with zero considering the uncertainties.
For D$_2$ gas data the fit yielded $\epsilon_T=(559\pm10\pm6) \times \,\mathrm{10^{-6}~K^{-1}}$, 
with very high significance in terms of uncertainty.

The central plots display the residuals for both model fits. 
No distinction between the models is visible in the Ne case, 
hence the parameter $\epsilon_T$ does not contribute.
In the D$_2$ gas case, this is clearly different: 
the LETA model shows systematic deviations that vanish
when using the YK model.
%
%Also in the second row of the figures, the residuals for both models look almost identical for Ne but reveal important differences for D$_2$.
The same is true in the histogram of residuals 
in the %third row of 
bottom plots in Figs.~\ref{fig:D2trans} 
and~\ref{fig:Netrans}.
%Almost no difference for Ne data but severe difference for the D$_2$ fitted with both models.
%The histogram for D$_2$ data fitted with LETA is not a good representation of the statistical fluctuations (e.g. two maxima), but fitted with the YK-model, the shape becomes a good representation of a statistical distribution.

The fit results are summarized in Table~\ref{tab:ned2comp}. 
As in the results for other gases (Table~\ref{tab:screening}), 
the uncertainties of $\epsilon_0$ are for both gases, D$_2$ and Ne, dominated by the systematic contribution 
from $N_\mathrm{dwell}$, whereas, 
for $\epsilon_T$ the uncertainties are dominated by statistics, as the latter is a relative parameter.
The neutron counts per measurement point for both detectors were above $3\times 10^5$.
However, considering only pure neutron count rate statistics would neglect other effects, such as minor changes in the neutron spectrum due to slow changes in the neutron production efficiency of the source.

The cross-check with Ne gas gives further confidence that in our current investigation, 
the function $N_\mathrm{dwell}$ fitted to the extended tail, 
which represents stored UCNs, is appropriate, even if improvable for measuring absolute cross sections with higher accuracy.

\begin{table}[htbp]
  \centering
  \caption{Best model fit parameters for Ne and D$_2$ data 
	with an additional parameter for a linear temperature dependence of 
	the up-scattering cross section according to the YK model as shown in Eq.~\ref{equ:et} 
	and Fig.~\ref{fig:Ratio-UCN-losses-upscattering-YK-and-LEA}. %\textcolor {red} {CORRECTED SYST ERRORS}
  }
    \begin{ruledtabular}
    \begin{tabular}{lcccccc}
%    \hline
    Species & {$\epsilon_0$}                    & {stat.} & {syst.} & {$\epsilon_T$} & {stat.} & {syst.} \\ % & {effective $\epsilon$} \\
%            & {($  \frac{\mu mol K^{1/2}}{m} &   $\pm$    & $\pm$      & {$10^{-6}/K$} &   $\pm$    & $\pm$      & ($ \frac{\mu mol K^{1/2}}{m}$) \\
            & $(\frac{\text{barn } m/s}{K^{1/2}})$ &   $\pm$    & $\pm$      & {$10^{-6}/K$} &   $\pm$    & $\pm$   \\ %    & $(\frac{\text{barn } m/s}{K^{1/2}})$ \\
   \hline
%    Ne    & 6060  & 60    & 600   & 54    & 38    & 10    & 5840 \\
%    Ne    & 100.6 &	1.0	& 10.0   & 54    & 38    & 10  \\ %  & 97.0 \\
   Ne    & 100.6 &	1.0	& 11.0   & 54    & 38    & 10  \\ %  & 97.0 \\
%    D$_2$    & 74600 & 230   & 3300  & 559   & 10    & 6     & 60600 \\
%    D$_2$    & 1238.8 &	3.8	& 54.8  & 559   & 10    & 6  \\ %   & 1006.3 \\
%    D$_2$    & 1238.8 &	37	& 66  & 559   & 20    & 18  \\ %   & 1006.3 \\
    D$_2$    & 1238.8 &	3.8	& 66  & 559   & 10    & 18  \\ %   & 1006.3 \\
%   \hline
    \end{tabular}%
    \end{ruledtabular}
  \label{tab:ned2comp}%
\end{table}%

%%The best fit of the Ne data results in $\epsilon_T=54 \cdot \,\mathrm{10^{-6}~K^{-1}}$, 
%%which is
%%compatible with zero
%%considering the uncertainties.
%
%%For D$_2$ gas data, $\epsilon_T=559 \cdot \,\mathrm{10^{-6}~K^{-1}}$, 
%%which is one order of magnitude higher than the $T$-dependence of Ne data but 
%%with significantly smaller uncertainties.

We interpret the large $\epsilon_T$ value in the case of D$_2$ as a clear demonstration of an additional $T$-dependent term in the YK-model 
compared to the LETA model for monoatomic gases.
Comparing this result to the theoretical prediction $\epsilon_T=(626\pm37) \times \,\mathrm{10^{-6} ~K^{-1}}$,  
and considering the measurement uncertainties and the unevenly distributed residuals (center plots in Figs.~\ref{fig:D2trans} 
and~\ref{fig:Netrans}) 
%would hint that there is some tension between measurement and theory.
%However, we would not argue into that direction before having deeper knowledge of 
hints towards yet unresolved systematic uncertainties in the measurements.
%, and possibly more accurate calculations of the theoretical model.

\begin{figure}
    \centering
    \includegraphics[width=1\linewidth]{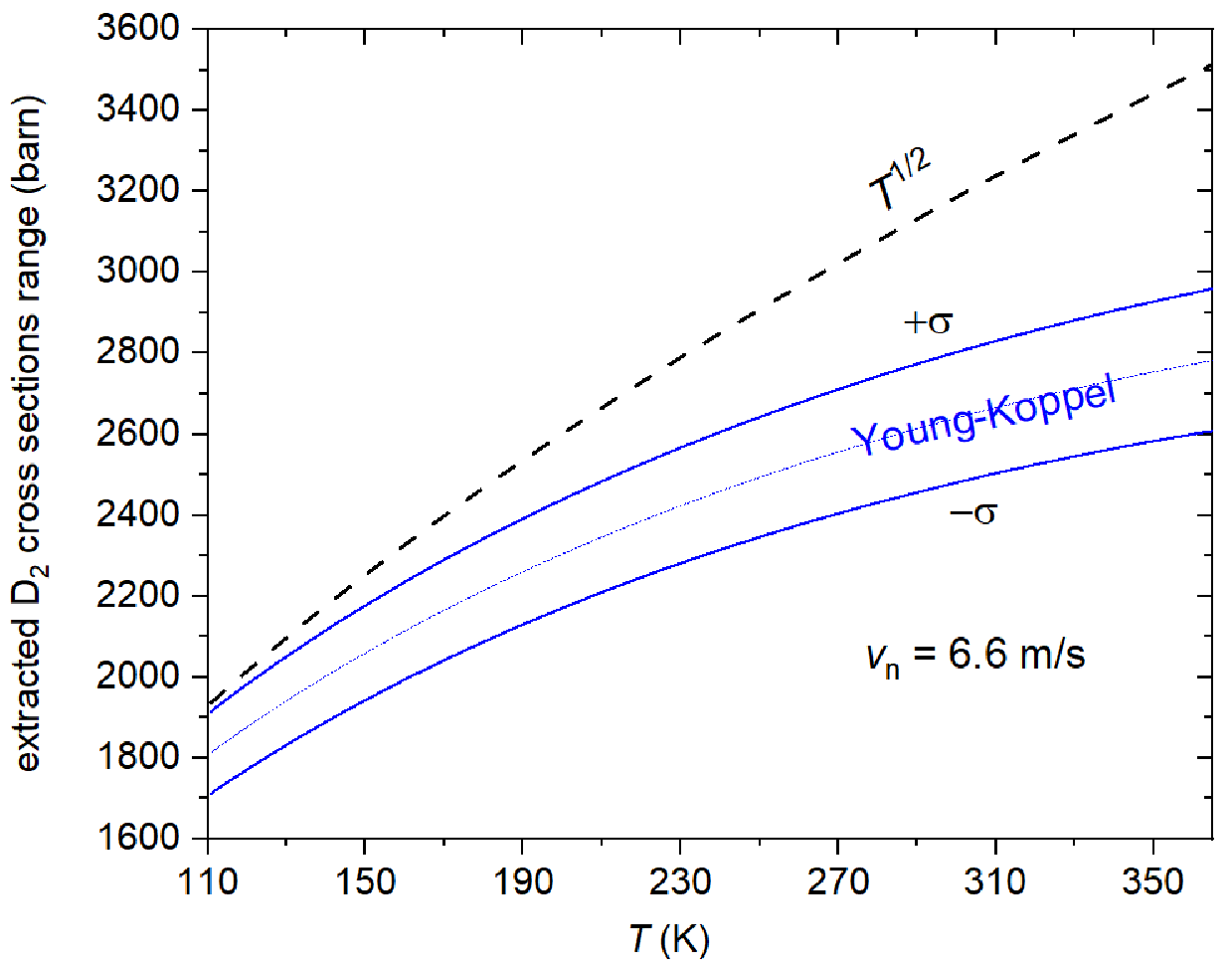}
    \caption{
%\textcolor {red} {
Temperature dependence of the D$_2$ total cross section applying the YK model to all measured data. To emphasize the difference, the dashed line indicates the  $\sqrt{T}$ dependence of the monoatomic approximation.%}
}
    \label{fig:ExtractedD2crosssectionVStemperature6.6mps}
\end{figure}

%\textcolor{red}{
Returning to Table~\ref{tab:screening}, the absolute value 
of the D$_2$ total cross section for UCNs is given under `literature' as calculated by integrating 
Eqs.~(A4) and (A5) in Ref.~\cite{YoungKoppel1964} and adding in a 2:1 ortho-para RT-equilibrium ratio. 
%\textcolor {red} {
Our measured value for RT is smaller than reported in~\cite{SeestromEtAl2015upscattering} by a factor of 1.2 of the quadratically combined error bars.
%}
%\textcolor{red}{
In Fig.~\ref{fig:ExtractedD2crosssectionVStemperature6.6mps} we plotted the D$_2$ total cross section as a function of temperature extracted by applying the YK model to the measured data using the parameters in Table~\ref{tab:ned2comp}. 
%}
%} 
%\textcolor{red}{
We also measured the total cross section of UCNs in H$_2$ gas. 
The data were analyzed by applying the Young-Koppel model, 
however, in a more limited temperature range (250-350\,K) 
and thus with less sensitivity to the details of this model. 
The theoretical YK differential cross sections, Eqs.~(29)-(30) in Ref.~\cite{YoungKoppel1964}, 
for para- and ortho-hydrogen were integrated over final energy and solid angle, and added in 
a 1:3 para to ortho equilibrium ratio at RT. 
The measurement result for H$_2$ and the theoretical expectation are in very good agreement, see Table~\ref{tab:screening}, consistent within the systematic uncertainties.
%\textcolor {red} {
Our extracted value for RT is smaller than reported in~\cite{SeestromEtAl2017upscattering} by a factor of 1.1 of the combined error bars.
%}
%}

\section{Summary and conclusion}
\label{sec:summary}

%--> Robin

%...Consequently, we reached our main goal, to provide a first experimental test over a wide temperature range of the YK-model vs LETA-models with high precision.

We performed UCN transmission measurements in D$_2$ and other gases at
temperatures between 110\,K and 365\,K. 
A comparison between the molecular deuterium and the monoatomic neon
gas showed a clearly different temperature dependence
of the total cross section. 
We analyzed our data using the Young-Koppel model 
and a Low-Energy Low-Temperature Approximation.
We interpret our best fit 
to the D$_2$ gas data
as a strong confirmation
for an additional temperature-dependent term 
due to spin correlations and rotational molecular states 
as described 
by the Young-Koppel model.
This is an important confirmation of the YK theory
and has implications for the understanding of 
deuterium-based ultracold neutron sources.

%\section{Acknowledgments}

%Swiss National Science Foundation, Michi, PSI groups

%KIT support 

\begin{acknowledgments}
We acknowledge the PSI proton accelerator operations section,
many PSI support groups, and 
the BSQ group for the excellent operation of the UCN source.
Excellent technical support by M. Meier is acknowledged. 
This work was supported by the Swiss National Science Foundation Project
178951, %Ingo's Gehalt ! A novel neutronics model of the solid D2 ultracold neutron converter at PSI aiming for increasing and 
and
IZSEZ0\_187982. %(Travel Exchange)
\end{acknowledgments}

%\section{References}

\bibliographystyle{apsrev4-2} % Tell bibtex which bibliography style to use

%\bibliography{UCN-references,YoungKoppelModel,nedm-references}
\bibliography{UCN-references,YoungKoppelModel}

\end{document}